\documentclass[twoside,fleqn]{article}

\usepackage{espcrc2}
\usepackage{epsfig}
\usepackage{psfig}
\newcommand{\beqn} {\begin{equation}}
\newcommand{\eqn} {\end{equation}}

\newcommand{\AmS}{{\protect\the\textfont2
  A\kern-.1667em\lower.5ex\hbox{M}\kern-.125emS}}
\def\NT{N_\tau}
\def\nt{\ifmmode\NT\else$\NT$\fi}
\def\NS{N_\sigma}
\def\ns{\ifmmode\NS\else$\NS$\fi}

\def\NP{{ Nucl.\ Phys.\ }}

\def\EN{{\langle P \rangle }}
\def\ENQ{{\langle P^2 \rangle }}

\title{Finite size analysis of the pseudo specific heat in SU(2)
       gauge theory
        \thanks{We thank the Deutsche Forschungsgemeinschaft for 
                the support of the work under grant Pe 340/3-3.}
       }

\author{J. Engels and T. Scheideler\address{Physics Department, 
        University of Bielefeld, 
        P.O. Box 100131, 33501 Bielefeld, Germany}
        }
       
\begin{document}

\begin{abstract}
We investigate the pseudo specific heat of SU(2) gauge theory near the 
crossover point on $4^4$ to $16^4$ lattices. Several different methods 
are used to determine the specific heat. 
The curious finite size dependence of the peak maximum 
is explained from the interplay of the 
crossover phenomenon with the deconfinement
transition occurring due to the finite extension of the lattice. 
In this context we calculate the modulus of the lattice average of the 
Polyakov loop on symmetric lattices and compare it to the prediction from 
a random walk model. 
\end{abstract}

\maketitle

\section{INTRODUCTION}

The pseudo specific heat $C_V$ of $SU(2)$ gauge theory was already 
investigated in the beginning of Monte Carlo lattice studies.
It is known to have a peak near $\beta=4/g^2 \approx 2.2$, in the
crossover region between strong and weak coupling behaviour.
A first finite size analysis by Brower et al. \cite{Brow81} on $4^4-10^4$
lattices revealed a strange dependence of the peak on the volume
$V=(N_{\sigma}a)^4$ of the lattice. Here, $a$ is the lattice spacing
and $N_{\sigma}$ the number of points in each direction. The location 
of the peak shifts with increasing volume from smaller $\beta-$values
to larger and then to smaller ones again; the peak maximum decreases
with increasing volume. Such a behaviour is unknown for any ordinary
phase transition. The nature and origin of the peak remained therefore
unclear, though a connection to the nearby endpoint of the first order
critical line in the ($\beta, \beta-$adjoint)-plane was proposed by
Bhanot and Creutz \cite{Bhan81}. Recently, new calculations for this 
extended $SU(2)$ model were performed \cite{Gava95,Step96}. Likewise, a new
study of $C_V$ with higher statistics and also on larger lattices,
utilizing the analysis techniques now available, seems appropriate.
In addition, since symmetric lattices are used to simulate zero 
temperature physics, it is important to estimate remaining finite
temperature effects which may be seen in the pseudo specific heat.

\section{METHODS TO CALCULATE $C_V$}

We use the standard Wilson action for $SU(2)$
\begin{equation}
S = \beta \cdot \sum_{x,\mu \nu} P_{\mu \nu}(x)~,
\end{equation}
where 
\begin{equation}
P_{\mu \nu}(x) = 1 - {1 \over 2}{\rm Tr}U_{\mu \nu}(x)~,
\end{equation}
is the plaquette or energy and $U_{\mu \nu}(x)$ is the plaquette
link operator. The sum extends over all independent forward plaquettes.
There are $N_P = 6N_{\sigma}^4$ such plaquettes. We denote the
lattice average of the plaquettes by $P$
\begin{equation}
P = {1 \over N_P} \sum_{x,\mu \nu} P_{\mu \nu}(x)~.
\end{equation}
The speudo specific heat is then defined by
\begin{equation}
C_V = { d\EN \over d(1/\beta) } = - \beta^2 { d\EN \over d\beta }~.
\end{equation}
There are three methods to determine $C_V$ :
\par \noindent
i) one measures the plaquette expectation values $\EN$ as a function
of $\beta$ and calculates the numerical derivative at $\beta_M=\beta+
\Delta\beta/2$ from
\beqn
C_V(\beta_M) = 
-{\beta_M^2 \over \Delta\beta} (\EN(\beta+\Delta\beta)
-\EN(\beta))~;
\eqn
ii) one measures the variance of the plaquettes, which is proportional
to $C_V$
\beqn
C_V = \beta^2 N_P (\ENQ-\EN^2)~;
\eqn
or,\par\noindent
iii) one calculates the sum of plaquette-plaquette correlations
\beqn
C_V = \beta^2 \sum_{x^{\prime},\mu^{\prime} \nu^{\prime}} (
\langle P_{\mu \nu}(x) P_{\mu^{\prime} \nu^{\prime}}(x^{\prime}) \rangle
-\EN^2)~.
\eqn
As it should be, all three methods are in complete consistence with each 
other. The most straightforward way is, of course, to calculate the 
variance of $\EN$. The density of states method (DSM) may then be
used to interpolate between the points. The variance becomes, however,
definitely too small, if the plaquettes are measured during and not
after each update, because that leads to local correlations among the 
plaquettes.
\begin{figure}[htb]
\begin{center}
   \epsfig{bbllx=161,bblly=301,bburx=414,bbury=551,
       file=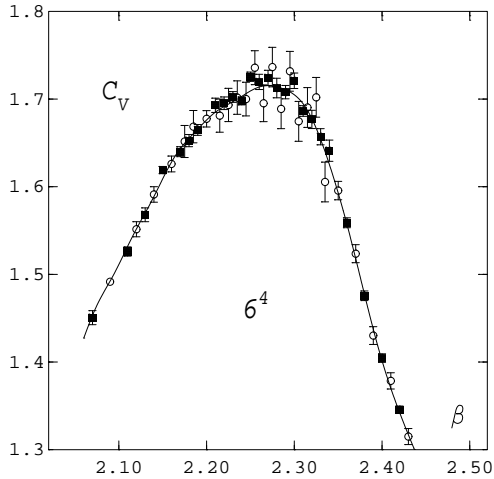, height=45mm}
\end{center}
\caption{The pseudo specific heat $C_V$ calculated from the variance
( squares), the numerical derivative (circles) and the DSM interpolation,
calculated on a $6^4$ lattice. }

\label{fig:cvn6}
\end{figure}
 
In Fig.~\ref{fig:cvn6} we show for comparison the results from methods 
i) and ii)
for a $6^4$ lattice. Here, on the average 90-120 thousand measurements
were made every fifth update at each $\beta-$value.
An update consisted of one heatbath and two overrelaxation steps.

We have also investigated the plaquette correlations. We find in general 
a rapid fall with $R=x^{\prime}-x$, the correlation length is of order 1.
The plaquettes $P_{\mu \nu}(x)$ and $P_{\mu^{\prime} \nu^{\prime}}(x^{\prime})$ 
may be in parallel or orthogonal planes. At the peak ($\beta \approx 2.23$)
we find that the total contribution of the orthogonal correlations is
about $30\%$ higher than that of the parallel correlations, whereas far away 
from the peak, at $\beta=2.70$, the contributions are essentially equal.

\section{FINITE SIZE DEPENDENCE OF $C_V$}

\begin{figure}[htb]
\begin{center}
    \epsfig{bbllx=140,bblly=250,bburx=495,bbury=610,
        file=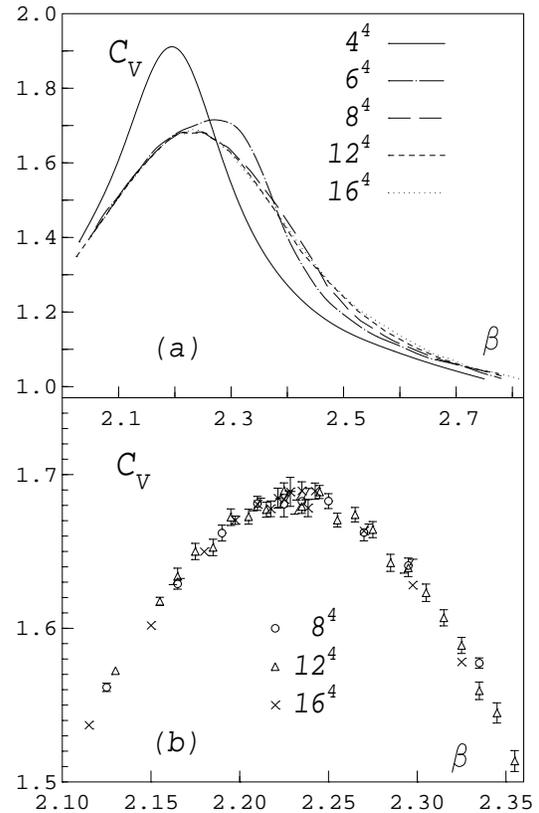, width=80mm, angle=-90}
\end{center}
\caption{The pseudo specific heat $C_V$ vs. $\beta$ on $\NS^4$ lattices. 
Part (a) shows the results for $\ns=4$ and 6 from the DSM interpolation,
for $\ns=8,12$ and 16 the measured points were connected by straight lines.   
Part (b) shows the peak region for $\ns=8,12$ and 16 in detail. } 

\label{fig:size}
\end{figure}

In Fig. \ref{fig:size} we compare the results for $C_V$ from lattices
with $\ns=4,6,8,12$ and 16. The general behaviour already found in
\cite{Brow81} is fully confirmed. On the other hand, we see that there
is no further finite size dependence in the peak region, if $\NS \geq 8$.
This suggests, that the finite size dependence 
of the smaller lattices is related to a 
\\[0.3cm]

\noindent
different phenomenon. Indeed, the critical point for the $\nt=4$ finite 
temperature deconfinement transition is at $\beta_c = 2.30$, very close to the
crossover peak positon. Since we are using periodic boundary conditions
for all directions, the approach to the critical point corresponding to
$\ns$ will influence the plaquette expectation values. To check this, we
have calculated the lattice average $L$ of the Polyakov loop $L(\vec{x})$
\beqn
L = {1\over N_\sigma^3} \sum_{\vec{x}} L(\vec{x})~;~
L(\vec{x}) =  {1\over 2}{\rm Tr}
\prod_{t=1}^{\NS} U_{ \vec{x},t }~~.
\eqn
    
\begin{figure}[htb]
\begin{center}
   \epsfig{bbllx=161,bblly=301,bburx=414,bbury=551,
       file=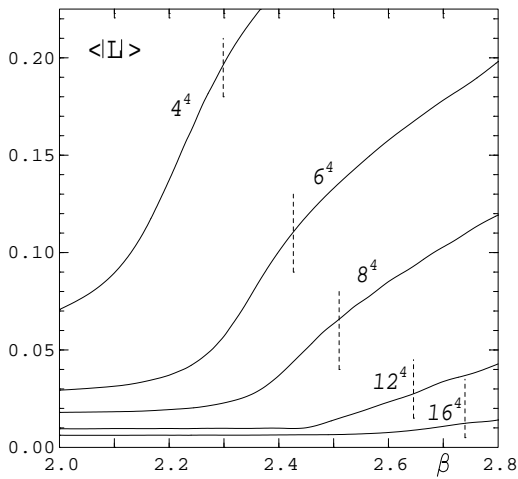, height=45mm}
\end{center}
\caption{The expectation value of the modulus of $L$ for $\NS=4,6,8,12$
and 16 vs. $\beta$. The broken vertical lines show the locations of the 
corresponding finite temperature phase transitions. }

\label{fig:Poly}
\end{figure}

As can be seen from Fig. \ref{fig:Poly}, the expectation value of the
modulus of $L$ is not zero on symmetric lattices, not even in the strong 
coupling limit, i.e. we have finite temperature effects also on symmetric
lattices. Well below $\beta_c(\NS)$ the quantity $\langle |L| \rangle$
is a constant. With increasing $\beta$ it starts to increase already before
the transition point. It is obvious, that due to the nearby transition
points the crossover peaks of the $4^4$ and $6^4$ lattices are stronger
distorted than those of the larger lattices, where only the right shoulders
of the peaks are slightly influenced. We may find the $\NS-$dependence
of $\langle |L| \rangle$ at $\beta=0$ from a simple random walk model, 
where the modulus of a  
     
\begin{figure}[htb]
\begin{center}
   \epsfig{bbllx=161,bblly=301,bburx=414,bbury=551,
       file=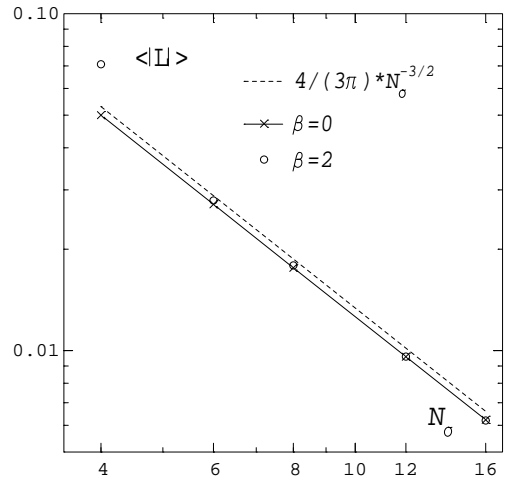, height=45mm}
\end{center}
\caption{The logarithm of $\langle |L| \rangle$ vs. ln$\ns$ on $4^4-16^4$
lattices at $\beta=0$ (solid line), at $\beta=2.0$ ( circles ) and from 
an estimate from the $\langle |L| \rangle$ value at $\ns=1$ (dashed line). }

\label{fig:logl}
\end{figure}

\noindent
sum over $\ns^3$ equal random variables is proportional
to $\ns^{3/2}$. Indeed, we obtain from a simulation at $\beta=0$ the   
following relation for $SU(2)$
\beqn
\langle |L| \rangle_{\beta=0} = 0.400 \cdot \ns^{-3/2}~~.
\eqn
\noindent
The $\langle |L| \rangle -$values, which we calculated at $\beta=2.0$  
are shown in Fig. \ref{fig:logl} together with the results for $\beta=0$.
At $\beta=2.0$ all lattices apart from the $4^4$ lattice have already 
reached the strong coupling value. Deviations
from this value indicate then the onset of finite temperature effects.

 Our final conclusion is, that the crossover peak is not
the result of an ordinary phase transition. For large lattices
the peak is at $\beta_{co}=2.23(2)$, its height is
$C_{V,co}=1.685(10)$. 

\end{document}